# Stable Xenon Nitride at High Pressures

Feng Peng[1,2], Yanchao Wang[1], Hui Wang[1], Yunwei Zhang[1], and Yanming Ma[1]*

[1]*State Key Lab of Superhard Materials, Jilin University, Changchun 130012, P. R. China*
[2]*College of Physics and Electronic Information, Luoyang Normal University, Luoyang 471022, P. R. China*
*Email of corresponding author: mym@jlu.edu.cn or mym@calypso.cn; webpage: http://mym.calypso.cn

Nitrides in many ways are fascinating since they often appear as superconductors, high energy density and hard materials. Though there exist a large variety of nitrides, noble gas nitrides are long missing in nature. Pursuit of noble gas nitrides has therefore become the subject of topical interests, but remains as a great challenge since molecular nitrogen ($N_2$, a major form of nitrogen) and noble gases are both inert systems and do not interact at normal conditions. We show through a swarm structure searching simulation that high pressure can lift the reactivity of both $N_2$ and xenon (Xe), making chemical reaction of them possible. The resultant nitride has a peculiar stoichiometry of $XeN_6$, possessing a high-energy-density of approximately 2.4 kJg$^{-1}$, rivaling that of the modern explosives. Chemically, $XeN_6$ is more intriguing with the appearance of chaired $N_6$ hexagons and an emergent 12-fold Xe by acceptance of unprecedentedly 12 Xe-N weak covalent bonds. Our work opens up the possibility of achieving Xe nitrides whose formation is long sought as impossible.

PACS numbers: 61.66.fn, 62.50.-p, 71.15.Mb, 81.40.Vw

Nitrogen (N) is the most abundant element in the Earth's atmosphere and is one of most important constituents of our universe. In air, N exists in the form of diatomic $N_2$ molecules and has the strongest-known triple N≡N bond. As a result, $N_2$ is chemically inert[1] and hardly interacts with other elements or substances under normal conditions. Syntheses of useful nitrides for industrial applications rely on chemical methods[2] via, e.g., temperature-programmed reaction, thermal decomposition, electrochemical synthesis, or with the aid of catalysts (e.g., the synthesis of $NH_3$).[3] Nitrides have a variety of intriguing properties, such as superconductivity[4], high energy density[5], and high hardness[6], as well as extraordinary chemical and thermal stability[7].

Noble gases (e.g., Ar, Kr, and Xe) as typical closed-shell systems are inert prototype examples and where the famous octet rule[8] originated. They are critical elements[9] because their abundances constrain the models for giant planet formation and the origin of their atmospheres. Among noble gases, Xe has the largest atomic core and is polarizable to form Xe fluorides[10] (or Xe noble metal hexafluorides[11]) and Xe oxides[12] in certain circumstances at ambient conditions due to its low ionization energy and large relativistic effect.[13] However, other Xe compounds are only synthesizable by using Xe fluoride precursors, instead of Xe gas, in the reaction with other organics or inorganics.

Since both $N_2$ and Xe are inert, they do not directly interact at ambient conditions. Although Xe and N containing compounds have been seen at ambient pressure in $(FXe)[N(SO_2F)_2]$ salts[14], the syntheses of these salts relied on the precursors of the cationic $(FXe)^+$ and anionic N-containing ligands, while not the uses of $N_2$ and Xe gases [13]. All of these as-synthesized complex salts have poor thermal stability, and most of them decompose explosively at room temperature.

The chemical reactivity of elements can be fundamentally modified under pressure[15]. As such, $N_2$ becomes reactive with noble metals to form technically important noble metal nitrides (e.g., $PtN_2$)[16] at moderate pressures. In these nitrides, N adopts a bonding pattern of the well-known singly bonded $N_2$ dimer[17]. Reactions of $N_2$ with CO are also possible in which the resultant products contain the singly bonded polymeric N and are potentially useful as high energy materials[5]. Reactivity of Xe can also be promoted at high pressures. Although experimental Van der Waals solids of Xe-$H_2$[18] and Xe-$H_2O$[19] were not characterized by a chemical interaction, the formation of various ionic binary Xe solids via direct reaction of the constituting elements (e.g., Xe-O[12,20], Xe-Fe/Xe-Ni[21], and Xe-Mg compounds[22]) was predicted under high pressures.

Because $N_2$ and Xe are both typical inert systems, their direct reaction is not thought as possible, even at high pressures. Earlier attempts[23] to acquire singly bonded polymeric N compressed a mixture of noble gases and $N_2$ up to 40 GPa but found no signs of reaction. As expected, Van der Waals mixtures were formed and were orientationally disordered, showing distinct properties from those of the constituting components. As a whole, Xe nitrides are long missing, greatly impeding the understanding of their physical and chemical properties of such nitrides.

In this work, we show using elaborative swarm-structure searching calculations[24,25] that $N_2$ reacts steadily with Xe at megabar pressures (>146 GPa) to form high-energy-density Xe nitrides. What surprised us is the emergent unique stoichiometry of $XeN_6$ and its entirely covalent N-N and N-Xe bonding network. The appearance of chaired $N_6$ hexagons and an ultrahigh 12-fold Xe via the acceptance of 12 Xe-N weak covalent bonds makes $XeN_6$ a fascinating high-energy-density nitride example.

Our structure searching simulations are performed through the swarm-intelligence based CALYPSO method[24] enabling a global minimization of free energy (at 0 K, it reduces to enthalpy) surfaces merging *ab initio* total-energy calculations as implemented in the CALYPSO code[25]. The method is specially designed for global structural minimization unbiased by any known structural information, and has been benchmarked on various known systems [24].

In sampling the energy surfaces, CALYPSO generates a series of structures. Total energy calculations and geometrical optimization for these structures were performed in the framework of density functional theory within the Perdew-Burke-Ernzerhof[26] parameterization of generalized gradient approximation[27] as implemented in the VASP (Vienna Ab Initio simulation package) code[28]. The projector-augmented wave (PAW) method[29] was adopted with the PAW potentials taken from the VASP library where $4d^{10}5s^25p^6$ and $2s^22p^3$ are treated as valence electrons for Xe and N atoms, respectively. The use of the plane-wave kinetic energy cutoff of 1000 eV and dense k-point sampling, adopted here, were shown to give excellent convergence of total energies.

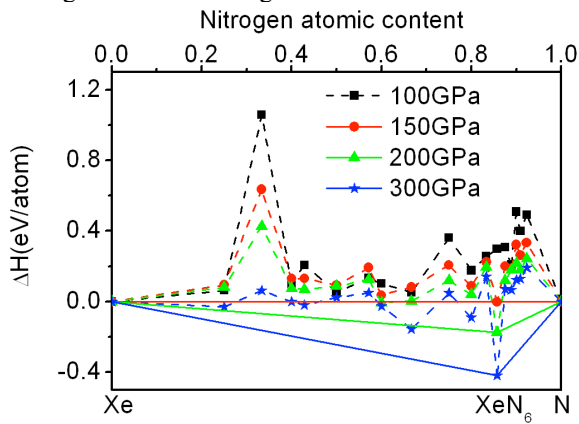

**Figure 1. Phase stabilities of Xe-N compounds.** The enthalpies of the formation of various Xe-N compounds under a series of pressures. The dotted lines connect the data points, and the solid lines denote the convex hull. The data show that $XeN_6$ is the only stable stoichiometry.

The phase stabilities of Xe-N system with various stoichiometries of $XeN_x$ ($x$ = 1/3, 1/2, 2/3, 3/4, 1, 4/3, 3/2, 2-10 & 12) were extensively investigated using calculations of the formation enthalpies[30] at pressures of 100, 150, 200, and 300 GPa, relative to the products of dissociation into constituent elements, as summarized in Fig. 1. The crystal structures used for enthalpy calculations were obtained from our first-principles swarm-structure searching simulations using CALYPSO code[24,25]. At 100 GPa, the formation enthalpies of all stoichiometries are positive, excluding $N_2$ from any reaction with Xe below 100 GPa. A stable phase emerges at 150 GPa, and $XeN_6$ appears to be the most stable and the only stable stoichiometry. Note that all the energy calculations in Fig. 1 were performed at 0 K, and inclusion of temperature further promotes the formation of $XeN_6$ compound (see blow P-T phase diagram).

By thoroughly searching the structures through CALYPSO at 150 GPa, it is found that $XeN_6$ stabilizes in a hexagonal structure (space group *R*-3*m*, three formula units per cell, Fig. 2a) with Xe and N atoms occupying 3*a* (0, 0, 0) and 18*h* (0.124, -0.124, -0.437), respectively. This hexagonal structure remains energetically most stable up to 300 GPa the highest pressure studied. Within the structure, a sublattice of N atoms adopts a chair-like form of hexameric $N_6$, structurally identical to the high-pressure phase of ε-sulfur (i.e., the $S_6$-ring structure)[31]. Intriguingly, the Xe sublattice is isomorphous with the high-pressure β-Po phase of tellurium (Te)[32]. The $XeN_6$ phase can thus be viewed as interpenetration of two sublattices of ε-phase of S and β-Po phase of Te.

At 150 GPa, the shortest Xe-Xe distance in the $XeN_6$ structure is 3.62 Å, much larger than twice the covalent radius (1.4 Å) of Xe[33], excluding any Xe-Xe bonding possibility. Each Xe has six nearest and six next-nearest neighboring N with Xe-N distances of 2.14 Å and 2.26 Å, respectively. These Xe-N distances are in the same distance range as (2.20 Å) in some salts[34] and (2.15 Å) of the sum of covalent radii of Xe and N atoms. As we will show below, each Xe forms 12 Xe-N bonds with those neighboring N. Each N has four neighbors of two N and two Xe. The nearest N-N distance is 1.35 Å, nearly identical to that (1.346 Å) of single N-N bond in cg-N at 115 GPa[35], but larger than the double N=N (1.20 Å) and triple N≡N (1.09 Å) bonds[36]. This result verifies the singly N-N bonding nature of the structure.

The electron localization function (ELF)[37] of $XeN_6$ was calculated for bonding analysis (Fig. 2b) because ELF provides a useful measure of the electron localization. The results showed that each Xe covalently bonds with 12 N atoms (Fig. 2d), while each N in $sp^3$ hybridization forms four covalent bonds with two Xe and two N atoms (Fig. 2c). The bonding results are further confirmed using calculations of the difference charge density (Supplemental Material Fig. S2). Note that the Xe-N bonds are strongly polarized towards N, possibly owing to the large difference in the atomic polarizabilities of N (7.6 a.u.) and Xe (27.8 a.u.). The Xe-N covalent bonding arises from the strong orbital hybridization between N-2*p* and Xe-5*p* as indicated by the projected density of states (PDOS) in Fig. 3b. Subsequently, electron topological analysis of the electron density was performed through Bader's quantum theory of atoms-in-molecules[38], which has been successfully applied into the determination of bonding interactions through the values of the density and its Laplacian at bond critical points. The calculated data are summarized in Table 1. Again, the analysis gives a strong support on the formation of N-N and N-Xe covalent bonds as indicted by the negative Laplacian values and large electron density at the critical points (Fig. S3).The bonding results are in excellent agreement with those derived from ELF analysis. While, the small Laplacian values at BCP of N-Xe indicate the weak covalent bonds.

**Table 1.** Bond critical point data of XeN$_6$ at 150 GPa.

| Bond | Position | $\nabla^2\rho(r_{cp})$ | $\rho(r_{cp})$ |
|---|---|---|---|
| N-N | 0.5000, 0.6855, 0.3145 | **-0.673** | 0.373 |
| N-Xe1 | 0.6339, 0.8342, 0.8342 | **-0.269** | 0.124 |
| N-Xe2 | 0.1682, 0.1682, 0.8418 | **-0.086** | 0.118 |

Note: $\rho(r_{cp})$ and $\nabla^2\rho(r_{cp})$ are the charge density (eÅ$^{-3}$) and its Laplacian at the corresponding critical points, respectively. The high charge densities and negative Laplacians of charge densities indicate the formation of a covalent bond.

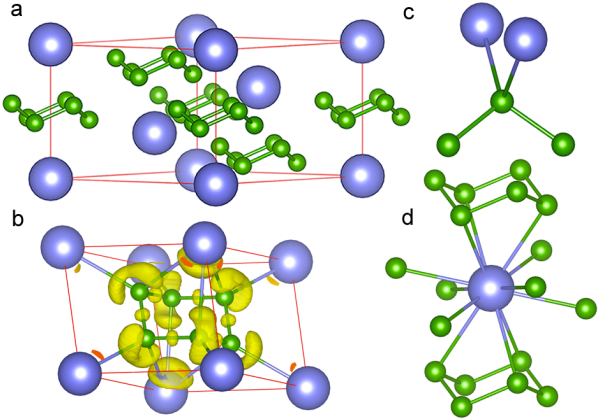

**Figure 2. Crystal structure and bonding properties of XeN$_6$.** (a) XeN$_6$ in an *R-3m* structure without showing any Xe-N bonds at 150 GPa where chaired N$_6$ hexagons are observed. The small green and large purple balls represent N and Xe atoms, respectively. (b) ELF plots at 150 GPa with an isosurface value of 0.83, showing covalent Xe-N bonding polarized towards N. (c) and (d) depict two structural units of 4-fold bonded N (i.e., N in $sp^3$ hybridization) and 12-fold bonded Xe, respectively.

If Xe retains a closed electron shell, it would not be possible to form any Xe-N covalent bonds. We show that the Xe-N bonding stems from the pressure-induced Xe→N charge transfer. A model system of hypothetical XeN$_0$ in which all N atoms were removed from the structure was constructed for PDOS calculations (Fig. 3a). Once N was introduced into the lattice, the completely filled 5$p$ valence states of Xe were partially depleted to the unoccupied orbital, showing a clear charge transfer of Xe to N. Bader charge analysis[38] reveals a transferred charge of approximate 2.4e from Xe-5$p$ to six N-2$p$ orbitals, which is caused by the orbital hybridization of Xe-5$p$ and N-2$p$ (Fig.3b).

Charge depletion of Xe leads to the opening up of the closed 5$p$ electron shell and Xe electronically behaves like a chalcogenide element of Te. By contrast, acceptance of 0.4 electrons per N gives N an electronic similarity to the chalcogenide S. These charge renormalizations rationalize the actual structural pattern of XeN$_6$ featuring the interpenetration of ε-phase of S and β-Po phase of Te. In view of the Te-like Xe and the S-like N, the formation of stable Xe-N compounds is not completely unexpected. Previous studies indeed showed that Te reacts with S and Se[39]. However, the octagonal S$_8$ (or Se$_8$) ring molecules in the synthesized compounds (e.g., TeS$_7$ and Te$_{0.32}$Se$_{7.68}$) are fundamentally different from the chaired N$_6$ ring structure of XeN$_6$.

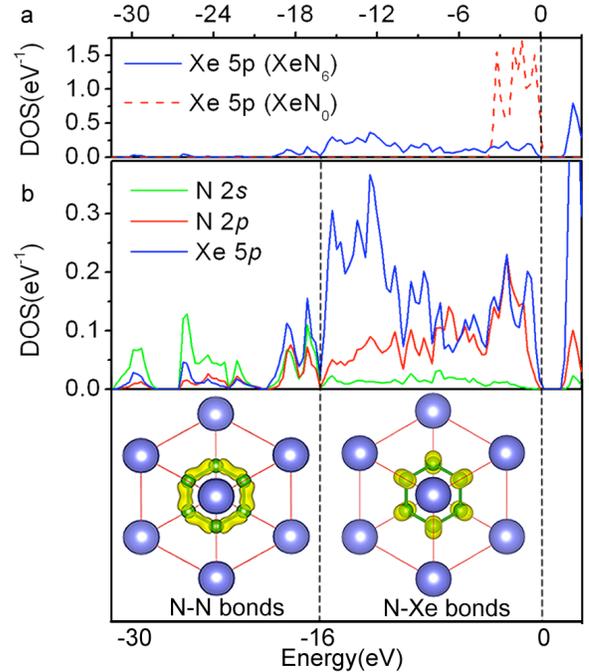

**Figure 3. Electronic PDOS and partial band decomposed charge density of XeN$_6$.** (a) PDOS of Xe-5$p$ states for XeN$_6$ and hypothetical XeN$_0$ at 150 GPa. The dashed line indicates the Fermi energy. (b) PDOS and partial band decomposed charge density of XeN$_6$ are shown in top and bottom panels, respectively. The right and left dashed lines denote the Fermi energy and the energy level of -16 eV, respectively. Partial band decomposed charge densities corresponding to energy windows of -32 to -16 eV and -16 to 0 eV are responsible for N-N and N-Xe covalent bonds, respectively. Hybrid functional Heyd-Scuseria-Ernzerhof (HSE06) for the electron exchange energy is used to better account for energy bandgap.

Lone pair electrons often appear in the N-$sp^3$ hybridization configuration of most known high-pressure nitrides and polymeric nitrogen, leading to 3-fold coordination of N. Here in XeN$_6$, lone pair electrons of N collapse with the formation of a fourth bond. Four-fold N-$sp^3$ hybridization has only been observed in light-elements nitrides (e.g., cubic BN, Si$_3$N$_4$, and C$_3$N$_4$)[6]. These light elements have close $sp$ orbital energies and easily form isotropic $sp$ hybridized bonds. Xe is different because its 5$s$ and 5$p$ orbital energies deviate largely. As a result, the $sp$ hybridization in Xe is not possible and only 5$p$ is involved in the bonding (Fig. 3b). The formed 4-fold N-$sp^3$ hybridization is uniquely anisotropic with the emergence of two stronger N-N and two weaker N-Xe covalent bonds as shown in Fig. 3b. Electron charges below -16 eV contribute to N-N bonding, while those above -16 eV are responsible for N-Xe bonding. The Xe→N transferred charge at 150 GPa is nearly identical to that of the B→N charge transfer (~ 0.4 e)[40] in cubic BN calculated at ambient pressure.

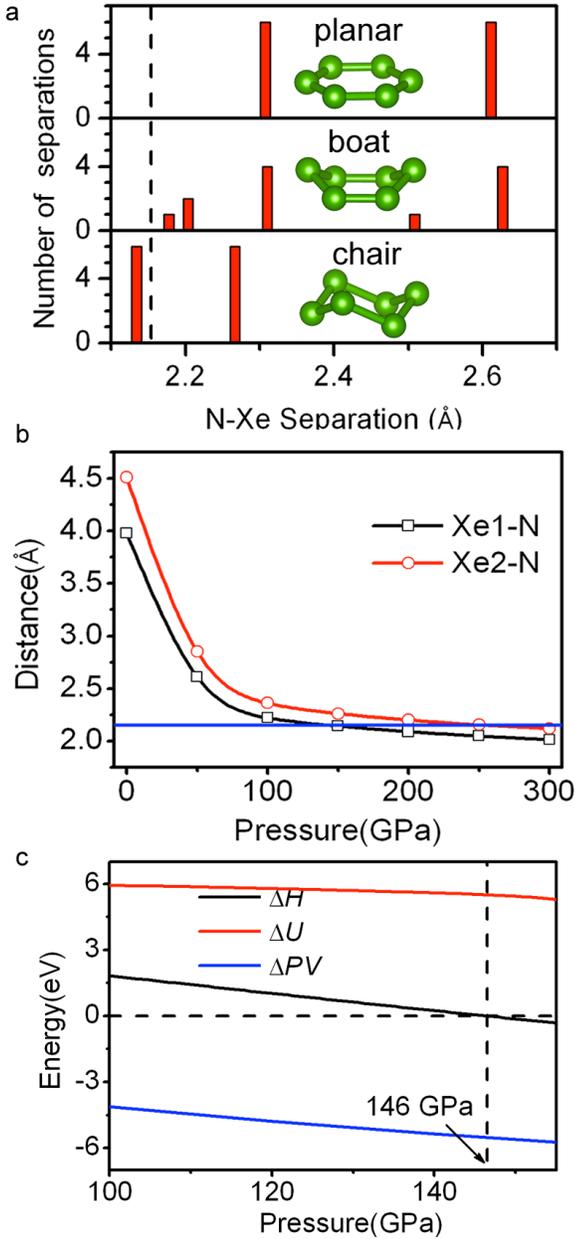

boat $N_6$ hexagons has one lone pair of electrons. We here purposely replaced our chaired $N_6$ hexagons with planar and boat ones in the $XeN_6$ structure. Using calculations of the total energies of these designed structures (Fig. 4a), it is observed that upon the formation of an extra Xe-N covalent bond after the sacrifice of the lone pair of electrons in the chaired $N_6$ hexagons, the system becomes energetically more preferred (Table 2). This simulation illustrates the great importance of the Xe-N covalent bonding in the stabilization of the $XeN_6$ nitride. It is worth noting that carbon rings in known cyclohexane molecules adopt similar chair forms where carbon is also in a 4-fold $sp^3$ hybridization [43].

**Table 2.** Enthalpies and volumes of $XeN_6$ with planar, chair, and boat $N_6$ units at 150 GPa.

| $N_6$ hexagons | Enthalpy(eV) | Volume (Å$^3$) |
| --- | --- | --- |
| planar | 13.07 | 46.65 |
| chair | 0.00 | 41.31 |
| boat | 17.28 | 46.30 |

Note: the enthalpies of $XeN_6$ with planar and boat $N_6$ hexagons are relative to the chair one. Formation of chair $N_6$ hexagon in $XeN_6$ is energetically most preferred.

We probe the formation mechanism of the Xe-N covalent bonds using calculations of the Xe-N distances with increasing pressure (Fig. 4b). The Xe-N distances decrease with pressure and reach the sum (2.15 Å) of the Xe-5$p$ and N-2$p$ covalent radii at 150 GPa, ready for the formation of an Xe-N bond. Indeed, analyses of the histograms of N-Xe separations (Fig. 4a) for chaired, planar, and boat $N_6$ hexagons in the designed $XeN_6$ nitrides revealed that the Xe-N separation that allowed for bonding can only be found in the chaired form.

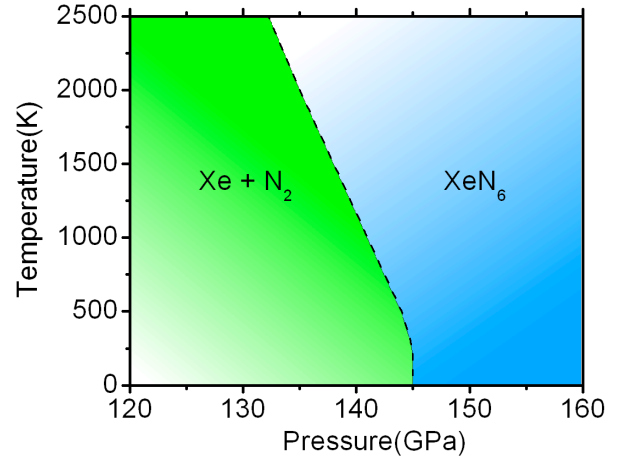

**Figure 4. Mechanism of pressure-driven reactivity of Xe with $N_2$.** (a) Histograms of N-Xe separations in $XeN_6$ with planar, chair, and boat $N_6$ units at 150 GPa. Dotted line presents the sum (2.15 Å) of Xe-5p and N-2p covalent radii. (b) Interatomic distances as a function of external pressures. Blue line presents the sum (2.15 Å) of Xe-5$p$ and N-2$p$ covalent radii. (c) Variation of $\Delta H$, $\Delta U$ and $\Delta PV$ with pressure for the formation of $XeN_6$. Gibbs free energy reduces to $H = U + PV$ at 0 K. For the chemical reaction of Xe + $3N_2 \rightarrow XeN_6$, we define $\Delta H = H_{XeN_6} - H_{Xe} - 3H_{N_2}$, $\Delta U = U_{XeN_6} - U_{Xe} - 3U_{N_2}$, and $\Delta PV = PV_{XeN_6} - PV_{Xe} - 3PV_{N_2}$, where $P$ and $V$ are pressure and volume per formula unit, respectively.

The chaired forms of $N_6$ hexagons in $XeN_6$ are in close correlation with the 4-fold N-$sp^3$ hybridization, which is in contrast to the planar $N_6$ rings in $LiN_3$.[41] A planar $N_6$ ring once subjected to a Jahn-Teller distortion deforms to a $C_{2v}$ 'boat' hexagon[42]. Note that each N in both planar and

**Figure 5. Phase diagram of Xe-$N_2$ system.** The dashed line separates the Xe-$N_2$ mixture (left and green region) and the stable $XeN_6$ compound (right and blue region).

Below, we discuss the underlying physical origin of the reaction of Xe and $N_2$ at high pressures using calculations of the enthalpies ($H=U+PV$), $PV$ terms, and static energies ($U$) of Xe, $N_2$ and $XeN_6$ as shown in Fig. 4c. The competition between $\Delta U$ and $\Delta PV$ determines the fate of the actual chemical reaction. Although $\Delta U$ remains largely positive, the $\Delta PV$ term is negative due to the volume contraction associated with a higher packing efficiency in XeN$_6$ and it is further reduced substantially as the pressure increases. At 146 GPa, the $\Delta PV$ term dominates the competition, leading to a negative $\Delta H$ value at which the stable XeN$_6$ compound forms.

We performed quasi-harmonic free-energy calculations to take account for the temperature effects with phonon spectra computed using the finite-displacement method[44]. We found that reactions of Xe and N$_2$ are significantly promoted upon heating. For example, at 146 GPa, the formation free energy is much preferred at -72 meV at 3000 K than -2 meV at 0 K (Supplemental Material Fig. S4). Figure 5 shows the computed pressure versus temperature phase diagram of XeN$_6$ with respect to the mixture of elemental Xe and N$_2$. Our results established that XeN$_6$ is steadily stable at pressure 132 GPa and a temperature of 2500 K (Fig. 5).

The predicted framework phase of XeN$_6$ with atomic forms of N is a potential high-energy-density material, since the three-dimensional N bonded phase is expected to decompose exothermically to solid Xe and molecular N$_2$ at ambient pressure. The chemical energy released is estimated to be 5.3 eV during this decomposition at the PBE level. This corresponds to an energy density of approximately 2.4 kJg$^{-1}$. The high energy content of XeN$_6$ rivals and exceeds that of the modern explosives such as TATB, RDX, and HMX, whose energy density values are in the range 1 to 3 kJg$^{-1}$.[45]

In summary, high pressure was applied to the mixture of N$_2$ and Xe through a swarm-structure searching simulation aiming to search for synthesizable Xe nitrides. The hitherto unknown high-energy-density XeN$_6$ nitride with a three-dimensional covalent bonding network was unraveled. The disclosed XeN$_6$ provides the opportunity for revealing the unexpected 12-fold Xe by acceptance of 12 weak polarized Xe-N bonds. We show the emergence of the Xe-N covalent bonding is crucial to stabilize the XeN$_6$ nitride. Our work opens up the possibility of achieving Xe nitrides whose formation is long thought as impossible. We also examined the reaction possibilities of N$_2$ with Kr, the upper neighbor of Xe, but found no signs of reaction up to 500 GPa (Supplemental Material Fig. S5). It is unlikely that other lighter noble gases (e.g., He, Ne, and Ar) with much smaller atomic cores can interact with N$_2$ below 500 GPa.

We thank the China 973 Program (2011CB808200), Natural Science Foundation of China under 11304141, 11304167, 11274136, 11104104, 11025418 and 91022029, the 2012 Changjiang Scholars Program of China, Changjiang Scholar and Innovative Research Team in University (IRT1132). This work is also sponsored by the Program for Science and Technology Innovation Research Team in University of Henan Province Grant No.13IRTSTHN020. Part of the calculations was performed in the High Performance Computing Center of Jilin University.

*Corresponding author.
mym@jlu.edu.cn or mym@calypso.cn